\documentclass{article}

\usepackage[creativeai,final]{neurips_2026}
\usepackage[utf8]{inputenc}
\usepackage[T1]{fontenc}
\usepackage{hyperref}
\usepackage{url}
\usepackage{microtype}
\usepackage{xcolor}
\usepackage{graphicx}
\usepackage{tikz}
\usetikzlibrary{positioning,arrows.meta}
\usepackage[normalem]{ulem} 

\definecolor{wcAudioFill}{RGB}{255,243,221}
\definecolor{wcAudioLine}{RGB}{196,120,30}
\definecolor{wcVisFill}{RGB}{228,237,247}
\definecolor{wcVisLine}{RGB}{66,105,155}
\definecolor{wcCtlFill}{RGB}{232,243,232}
\definecolor{wcCtlLine}{RGB}{74,130,84}

\newcommand{\wcic}[1]{\tikz[baseline=-0.55ex,x=1mm,y=1mm,line width=0.28pt]{#1}}
\newcommand{\icMic}{\wcic{\draw[rounded corners=0.35] (-0.45,0.55) rectangle (0.45,2.15);
  \draw (-0.95,1.05) arc (180:360:0.95);\draw (0,0.1)--(0,-0.5);}}
\newcommand{\icALM}{\wcic{\fill (-1,0.2) circle (0.28);\fill (0,1.6) circle (0.28);
  \fill (1,0.45) circle (0.28);\fill (0.05,-0.4) circle (0.28);
  \draw (-1,0.2)--(0,1.6)--(1,0.45)--(0.05,-0.4)--cycle;}}
\newcommand{\icSearch}{\wcic{\draw (0.05,1.0) circle (0.85);\draw (0.7,0.35)--(1.45,-0.4);}}
\newcommand{\icFilter}{\wcic{\draw (-1.2,1.9)--(1.2,1.9)--(0.28,0.6)--(0.28,-0.5)--(-0.28,-0.1)--(-0.28,0.6)--cycle;}}
\newcommand{\icGear}{\wcic{\draw (0,0.85) circle (0.8);\draw (0,1.65)--(0,2.1);
  \draw (0,0.05)--(0,-0.4);\draw (-0.8,0.85)--(-1.25,0.85);\draw (0.8,0.85)--(1.25,0.85);}}
\newcommand{\icSpeaker}{\wcic{\draw (-1.35,0.35) rectangle (-0.65,1.35);
  \draw (-0.65,0.35)--(0.1,-0.15)--(0.1,1.85)--(-0.65,1.35)--cycle;
  \draw (0.5,0.3) arc (-55:55:0.95);\draw (1.0,0.1) arc (-55:55:1.3);}}
\newcommand{\icGrid}{\wcic{\foreach \x in {0,1,2}{\foreach \y in {0,1,2}{%
  \fill (\x*0.72-0.72,\y*0.72+0.1) rectangle ++(0.48,0.48);}}}}
\newcommand{\icMap}{\wcic{\draw (-1.45,0.35) rectangle (-0.6,1.2);
  \draw (0.6,0.35) rectangle (1.45,1.2);\draw[-{Stealth[length=1mm]}] (-0.5,0.775)--(0.5,0.775);}}
\newcommand{\icCube}{\wcic{\draw (0,2.0)--(1.2,1.35)--(1.2,0.15)--(0,-0.5)--(-1.2,0.15)--(-1.2,1.35)--cycle;
  \draw (0,2.0)--(0,0.75);\draw (0,0.75)--(1.2,1.35);\draw (0,0.75)--(-1.2,1.35);}}
\newcommand{\icScreen}{\wcic{\draw (-1.3,0.5) rectangle (1.3,2.0);
  \draw (0,0.5)--(0,0.0);\draw (-0.7,0.0)--(0.7,0.0);}}
\newcommand{\icRails}{\wcic{\draw (-1.2,0.1) rectangle (1.2,1.9);\draw (-0.28,0.1)--(-0.28,1.9);}}

\title{Windchime: an Audiovisual Installation with Audio-Language Model implementations}

\author{%
  Trent Eriksen \\
  Independent Researcher \\
  San Francisco, USA \\
  \texttt{starduststereo@gmail.com} \\
  \And
  Celeste Betancur \\
  CCRMA, Stanford University \\
  Stanford, USA \\
  \texttt{celeste@ccrma.stanford.edu} \\
}

\begin{document}
\maketitle

\begin{abstract}
Windchime is an audiovisual installation in which spoken language retrieves music and interactive animations from an artist authored corpus. A local Automatic Speech Recognition (ASR) model transcribes each prompt while an audio-language model (ALM) embeds it and retrieval over 406 audio stems returns a soundscape with live coded \texttt{Strudel} fragments that layer into audio output. Simultaneously each spoken input results in retrieval of interactive \texttt{Three.js} animation scenes each containing unique tactile control via monome grid and arc hardware. Windchime distributes agency in three distinct directions; across the artist who authored the possible outputs, the ALM that influences what areas of the corpus can be retrieved, and the visitor whose spoken inputs result in multimodal embodied interaction.
\end{abstract}

\section{Description of the Work}
\label{sec:description}

The project's namesake is inspired by the tubular bells of a physical wind chime. Although these bells are fixed, when coming in contact with the wind, infinite outcomes are produced in the form of an evolving algorithmic melody. From this everyday object that exemplifies a physical algorithm, Windchime seeks to expand on this type of phenomenon where a corpus of artist authored audiovisual material becomes subject to emergent recombination based on spoken input.

\begin{figure}[t]
\centering
\includegraphics[height=44mm]{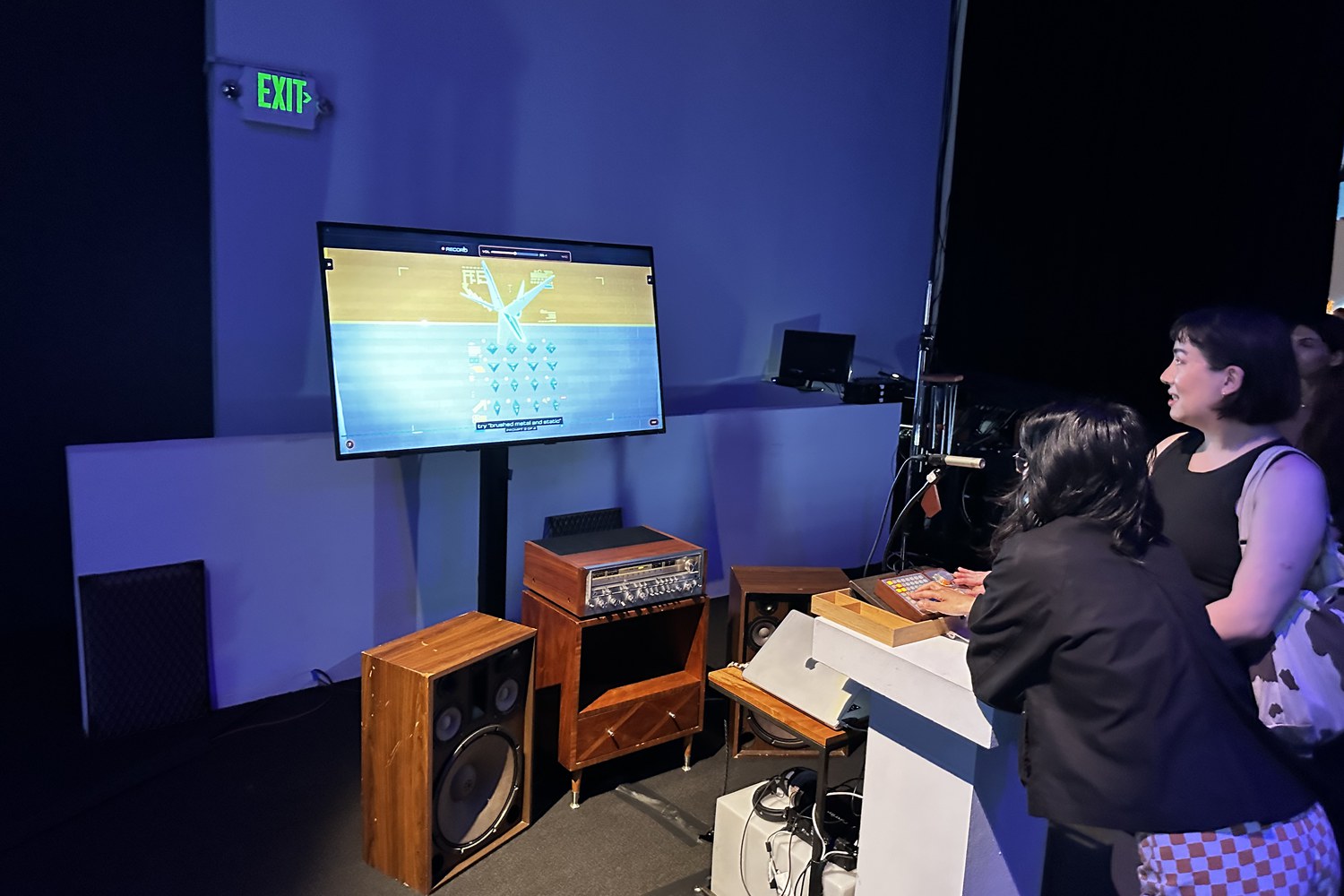}\hfill
\includegraphics[height=44mm]{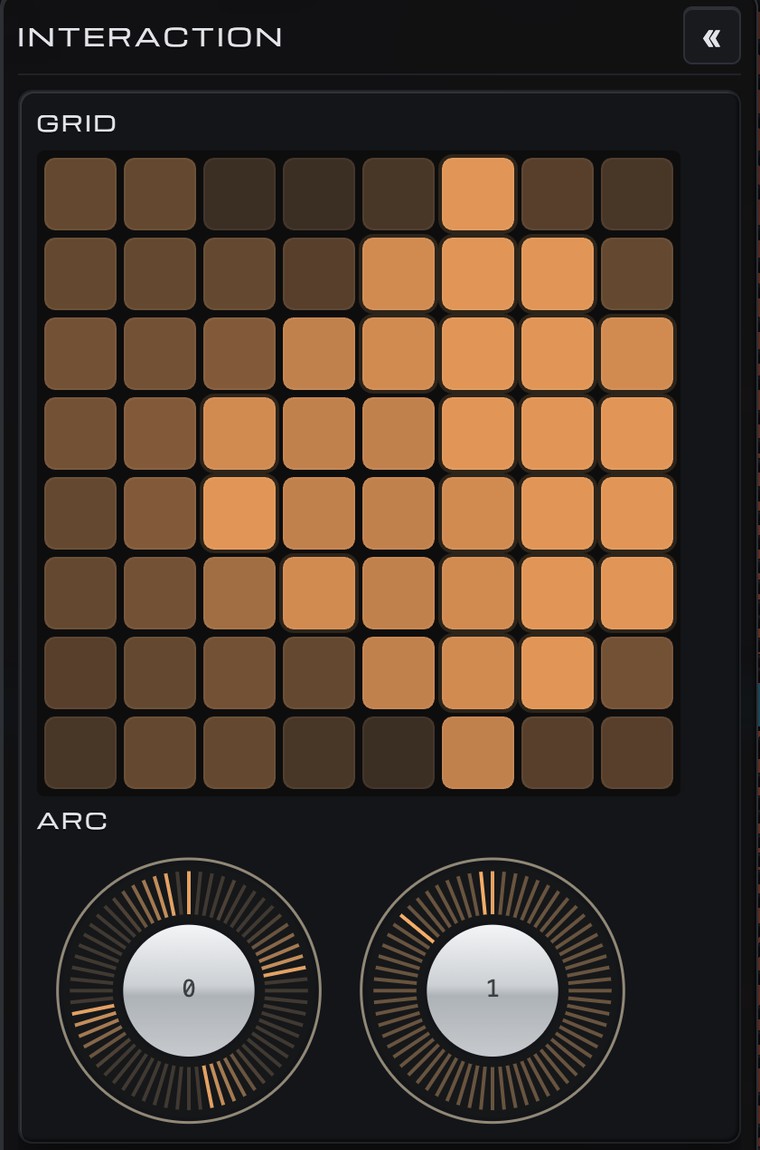}\hfill
\includegraphics[height=44mm]{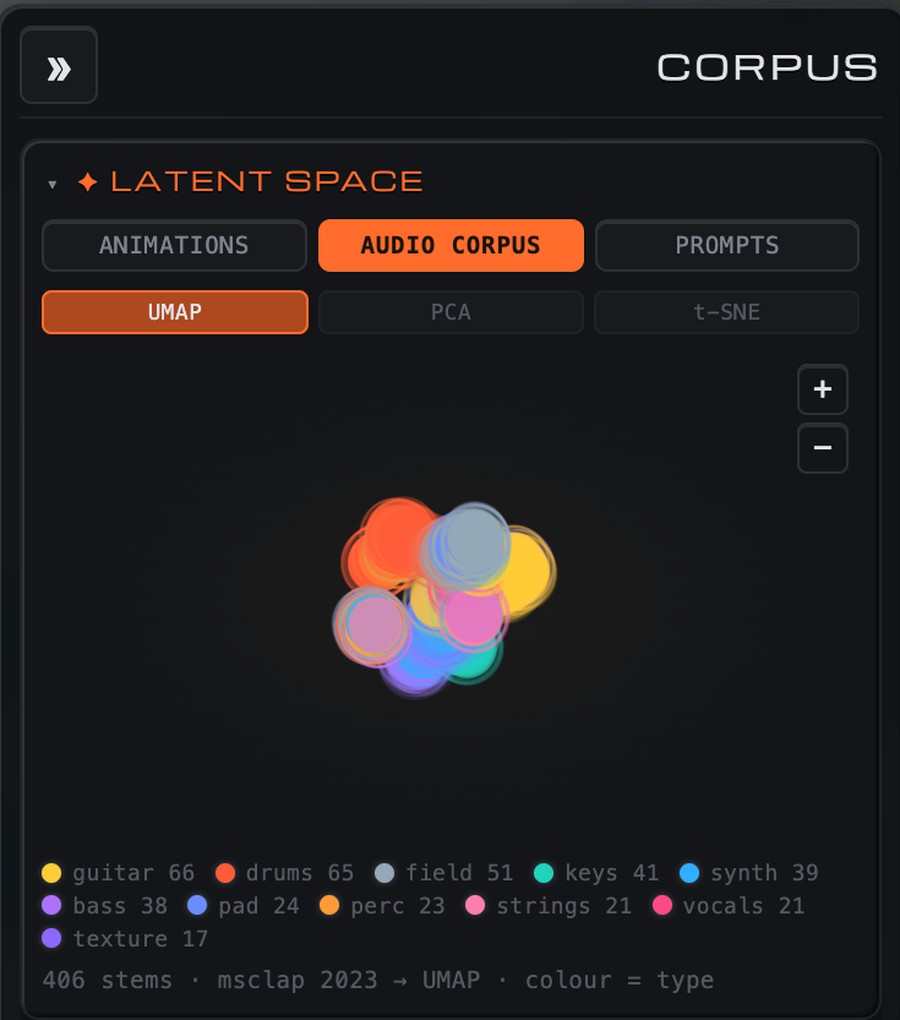}
\caption{Windchime at the Gray Area Cultural Incubator Showcase. On the left are two visitors with the installation. In the center the interaction rail mirrors the connected monome hardware as a digital twin. On the right, the corpus rail places the audio stems in the latent space of the active audio-language model, colored by instrument role.}
\label{fig:install}
\end{figure}

The installation comprises a screen based visual field, a microphone, monome grid and arc tactile controllers \citep{vallis2010iterative}, and a vintage hi-fi sound system (Figure~\ref{fig:install}). These components enable embodied interaction through voice, touch, vision and sound. The interaction design is organized into segments consisting of intro and outro sequences and four spoken inputs during which the visitor is free to control animations and explore the interface for further legibility. The four spoken inputs are limited to 15 seconds each and can range from descriptive to imagistic, performative and open phrasings. Each spoken input surfaces a unique combination of audio and controllable animations after which a summary of the interaction is shown with latent space animations and the installation resets for the next visitor. This interaction loop is designed to ensure accessibility to multiple visitors in a public venue. Windchime debuted at the \href{https://grayarea.org/event/cultural-incubator-showcase-2026/}{Gray Area Cultural Incubator Showcase} from August 5 to 7, 2026 in San Francisco, California with video demo available \href{https://youtu.be/j3XyW2ynH5Y}{here}. 

In the retrieval pipeline four sound modes called 'Peak', 'Soundscape', 'Immediate' and 'Settled' shape which audio file sections become audible. 'Peak' starts at the most energetic 30 seconds of the file, 'Soundscape' plays from the beginning, 'Immediate' from the first audible section and 'Settled' at the earliest 30 seconds with two-thirds of peak energy. Windchime is configurable with different sound modes on each prompt and the aforementioned order is what appeared in the public installation. This choice has a substantial effect as the audio files vary in length and content leading to distinct sonic outputs based on the combination of ALM, sound mode and live coded audio in \texttt{Strudel} \citep{roos2023strudel} a popular choice for music creation among creative coders and audiovisual artists.

The audio corpus consists of musical audio and field recordings from Trent Eriksen's projects. Musical audio examples contain stems, which are isolated recordings of instruments like guitars, bass, drums, percussion, orchestral strings, synthesizers, vocals and sound effects. Field recordings were captured onsite in Japan, Korea, the European Union and the United States in both urban and rural settings, allowing for unique multicultural outputs. It has been shown that natural language descriptions of music range from the technical to the situational \citep{salganik2025musicsem}, making this heterogeneous corpus well suited to various types of prompts.

Additionally, the animation corpus consists of 135 \texttt{Three.js} scenes primarily of an abstract style inspired by experimental animation, absolute film and early computer graphics. The monome grid and arc are chosen for their distinct minimalist aesthetic and fully programmable interactivity which allows for distinct application control and LED feedback per animation. This customization uniquely enables tactile gestural variety as a defining component of Windchime's embodied interaction. The user interface contains two collapsible rails that expose additional system information for increased legibility, with the left containing a responsive digital twin of the monome hardware and on the right, interactive 3D multimodal embeddings for audio, animation and prompts latent spaces.

\section{Role of AI and ML}
\label{sec:role}

\begin{figure}[t]
\centering
\begin{tikzpicture}[
  x=1mm, y=1mm,
  abox/.style={draw=wcAudioLine, fill=wcAudioFill, thick, rounded corners=2pt,
               align=center, inner sep=1.2mm, minimum width=20mm, minimum height=12mm,
               font=\scriptsize},
  vbox/.style={draw=wcVisLine, fill=wcVisFill, thick, rounded corners=2pt,
               align=center, inner sep=1.2mm, minimum width=20mm, minimum height=12mm,
               font=\scriptsize},
  nbox/.style={draw=wcVisLine, fill=wcVisFill, thick, densely dashed, rounded corners=2pt,
               align=center, inner sep=1.2mm, minimum width=20mm, minimum height=12mm,
               font=\scriptsize},
  tab/.style={draw=wcCtlLine, fill=wcCtlFill, thick, rounded corners=1.5pt,
              align=center, inner sep=0.7mm, minimum width=20mm, minimum height=4.5mm,
              font=\tiny},
  flow/.style={-{Stealth[length=2mm]}, semithick},
  cond/.style={-{Stealth[length=2mm]}, semithick, densely dashed}
]
\node[abox] (asr)    at (10,0)    {\icMic\\speech\\ASR};
\node[abox] (alm)    at (33.5,0)  {\icALM\\ALM\\one of six};
\node[abox] (retr)   at (57,0)    {\icSearch\\retrieval\\406 stems};
\node[abox] (policy) at (80.5,0)  {\icFilter\\one stem\\per role};
\node[abox] (plan)   at (104,0)   {\icGear\\ strudel planner};
\node[abox] (snd)    at (127.5,0) {\icSpeaker\\soundscape};
\node[tab, anchor=north] at (alm.south)    {ALM choice};
\node[tab, anchor=north] at (retr.south)   {sound mode};
\node[tab, anchor=north] at (policy.south) {role folders};
\node[vbox] (mon)  at (10,-22)   {\icGrid\\monome\\grid and arc};
\node[vbox] (map)  at (33.5,-22) {\icMap\\scene\\parameters};
\node[vbox] (fam)  at (57,-22)   {\icCube\\Three.js\\one of 135};
\node[vbox] (proj) at (80.5,-22) {\icScreen\\visual output};
\node[nbox] (rail) at (104,-22)  {\icRails\\ corpus rails};
\draw[flow, wcAudioLine] (asr) -- (alm);
\draw[flow, wcAudioLine] (alm) -- (retr);
\draw[flow, wcAudioLine] (retr) -- (policy);
\draw[flow, wcAudioLine] (policy) -- (plan);
\draw[flow, wcAudioLine] (plan) -- (snd);
\draw[flow, wcVisLine] (mon) -- (map);
\draw[flow, wcVisLine] (map) -- (fam);
\draw[flow, wcVisLine] (fam) -- (proj);
\draw[cond, wcVisLine] (snd.south) -- ++(0,-7.25)
  -| node[pos=0.32, fill=white, inner sep=0.6mm, font=\tiny]{optional coupling via monome takeover} (fam.north);
\end{tikzpicture}
\caption{Amber is the audio path from speech to soundscape, the three green tabs are the author's choices and constraints, while blue is the tactile and visual path controlled by monome hardware. Audio features interact with animations through the dashed coupling the visitor enables. Dashed rails are the observation surfaces of Section~\ref{sec:description}.}
\label{fig:system}
\end{figure}
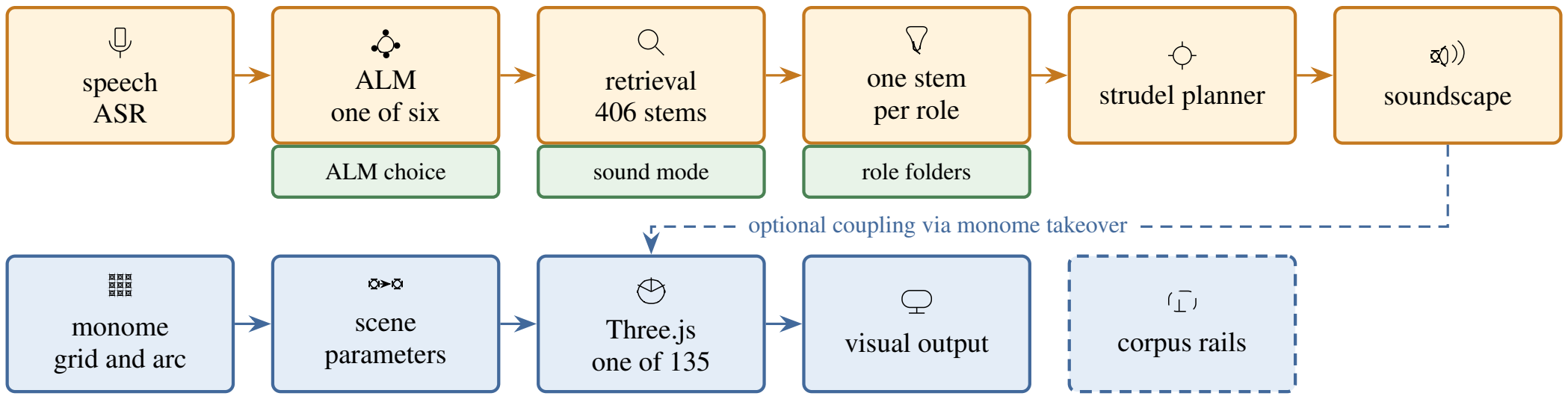

Machine Learning (ML) operates at three points in the system (Figure~\ref{fig:system}). A local faster-whisper model first transcribes the spoken prompt \citep{radford2023whisper}, then the selected ALM projects the transcript into a shared embedding space. The installation is configurable with multiple ALMs for example a landmark model based on Contrastive Language Audio Pretraining \citep{elizalde2023clap,wu2023laionclap}, and a model that supports multilingual input \citep{wu2025clamp3} among others. Cosine similarity over a FAISS index \citep{johnson2021faiss} of precomputed stem embeddings ranks the corpus and a selection policy admits at most one stem per instrument role among the forty nearest. Descriptors of the retrieved set follow, computed with librosa \citep{mcfee2015librosa} as energy, brightness, key, tempo, and onset density. A planner then reads them and chooses a \texttt{Strudel} template and its parameters to complement the retrieved audio. The planner queries a hosted large language model, falling back to a local model, then to a deterministic template, so the work degrades gracefully without network access and remains configurable to multiple settings.

The ALM is interchangeable while the corpus, index, similarity, and selection policy are fixed. The available ALMs are a general LAION-CLAP model as the default, a re-indexed copy of it, a checkpoint trained for music \citep{wu2023laionclap}, the 2022 and 2023 releases of MS-CLAP \citep{elizalde2023clap,elizalde2024nlsupervision}, and the multilingual CLaMP~3 \citep{wu2025clamp3}. This selection of models exemplifies a broad array of state of the art Music Information Retrieval (MIR) tools. While developing the installation it was noted that changing the ALM alters corpus retrieval making it a curatorial choice and a promising direction for further inquiry. The auditing of ML implementations in retrieval based artworks like this is a viable future research direction in the vein of other retrieval based creative tools \citep{brade2023synthscribe}.

\section{Addressing the Theme of Agency}
\label{sec:agency}

The choice of ALM conditions which parts of the audio corpus a prompt can reach as some models retrieve more distinct stems than others, giving it substantial influence on the experience of the installation. These models distribute agency away from both the artist and the visitor. The artist no longer selects what plays from their own material as the visitor's words set the query without fixing the outcome. The introduction of audio file directory categories and preset creation for animations act as artistic guardrails and limitations that constrain the installation to certain aesthetic bounds of audiovisual output. These kinds of choices reassert the artist's agency.

This distribution of agency has the potential to result in structured surprise as visitors may naturally contemplate the connection between their utterance and the installation's response. Windchime's audio retrieval architecture contains three elements that directly explore agency. The ALM that maps words to sounds, the sound mode that decides which parts of retrieved stems become audible, and the stem category folders the selection policy draws from. Some pairings of ALM and sound mode elicit more musical assemblages while others are frenetic, dynamic or ambient. Calibrating these components is an artistic task that displays how ML implementations can become compelling creative tools when embedded into interactive audiovisual installations.

In addition to voice input, the visitor's agency is sustained through tactile monome interaction mapped to animation control. This separation of language impacting audio output and tactility controlling animation provides a unique interaction design. Interactive AI art can draw every visitor toward one narrow region of a model's range \citep{qiu2023sameness} and Windchime explores this problem space directly by enabling ALM interchangeability. This practice facilitates interpretability via differing latent representations demonstrated by animation sequences and the user interface. For example, Windchime's UMAP in (Figure~\ref{fig:install}) can allow for artists and visitors working with experiential AI to understand the effects of different ML implementations on agency \citep{hemment2023agency}. In Windchime agency is intentionally distributed. The artist authors the corpus and sets the configuration. The visitor initiates every retrieval and the model settles the correspondence between spoken utterances and the artist's material. Supplemental material is provided \href{https://drive.google.com/drive/folders/1ELRImHqsBoD5yinlUPNVbSsZ8j_NPn58?usp=drive_link}{here}.

\section{Author Biography}
\label{sec:bio}

Trent Eriksen is an artist and researcher working across Music Information Retrieval, live coding, and audiovisual installations. His work has been shown across North America, Europe, and Asia with documentation collected at \url{https://grashopr-888.github.io/current-projects}.

Celeste Betancur is a PhD candidate at Stanford University's Center for Computer Research in Music and Acoustics whose portfolio is located at \url{https://www.celestebetancur.com/}.

\bibliographystyle{plainnat}
\bibliography{REVISION_REFERENCES}

\end{document}